\documentclass[12pt,halfline,a4paper]{ouparticle}
\pdfoutput=1
\usepackage{natbib}
\usepackage{siunitx}
\usepackage{booktabs}
\usepackage{textcomp}
\usepackage{multirow}
\usepackage{url}
\usepackage{graphicx}
\usepackage{listings}
\usepackage{courier}
\begin{document}
\title{Aliased seabed detection in fisheries acoustic data}
\author{
\name{Robert Blackwell}
\address{British Antarctic Survey,
  High Cross,
  Madingley Road,
  Cambridge,
  CB3 0ET,
  United Kingdom}
\email{roback28@bas.ac.uk\thanks{Corresponding author}}
\and
\name{Richard Harvey}
\address{University of East Anglia,
  School of Computing Sciences,
  Norwich Research Park, Norwich,
  NR4 7TJ,
  United Kingdom}
\email{r.w.harvey@uea.ac.uk}
\and
\name{Bastien Queste}
\address{University of East Anglia,
  School of Environmental Sciences,
  Norwich Research Park,
  Norwich,
  NR4 7TJ,
  United Kingdom}
\email{b.queste@uea.ac.uk}
\and
\name{Sophie Fielding}
\address{British Antarctic Survey,
  High Cross,
  Madingley Road,
  Cambridge,
  CB3 0ET,
  United Kingdom}
\email{sof@bas.ac.uk}}
\abstract{
  Aliased seabed echoes, also known as ``false bottoms'' or ``shadow
  bottoms'', are a form of echogram corruption caused by seabed
  reverberation from preceding pings coinciding with echoes from the
  current ping. These aliases are usually either avoided by adjusting
  the survey parameters, or identified and removed by hand - a
  subjective and laborious process.

  This paper describes a simple algorithm that uses volume
  backscatter and split-beam angle to detect and remove aliased seabed
  using single frequency, split-beam echo sounder data without the
  need for bathymetry.
}
\date{\today}
\maketitle
\section{Introduction}
\begin{figure*}
\centering \includegraphics[width=\textwidth]{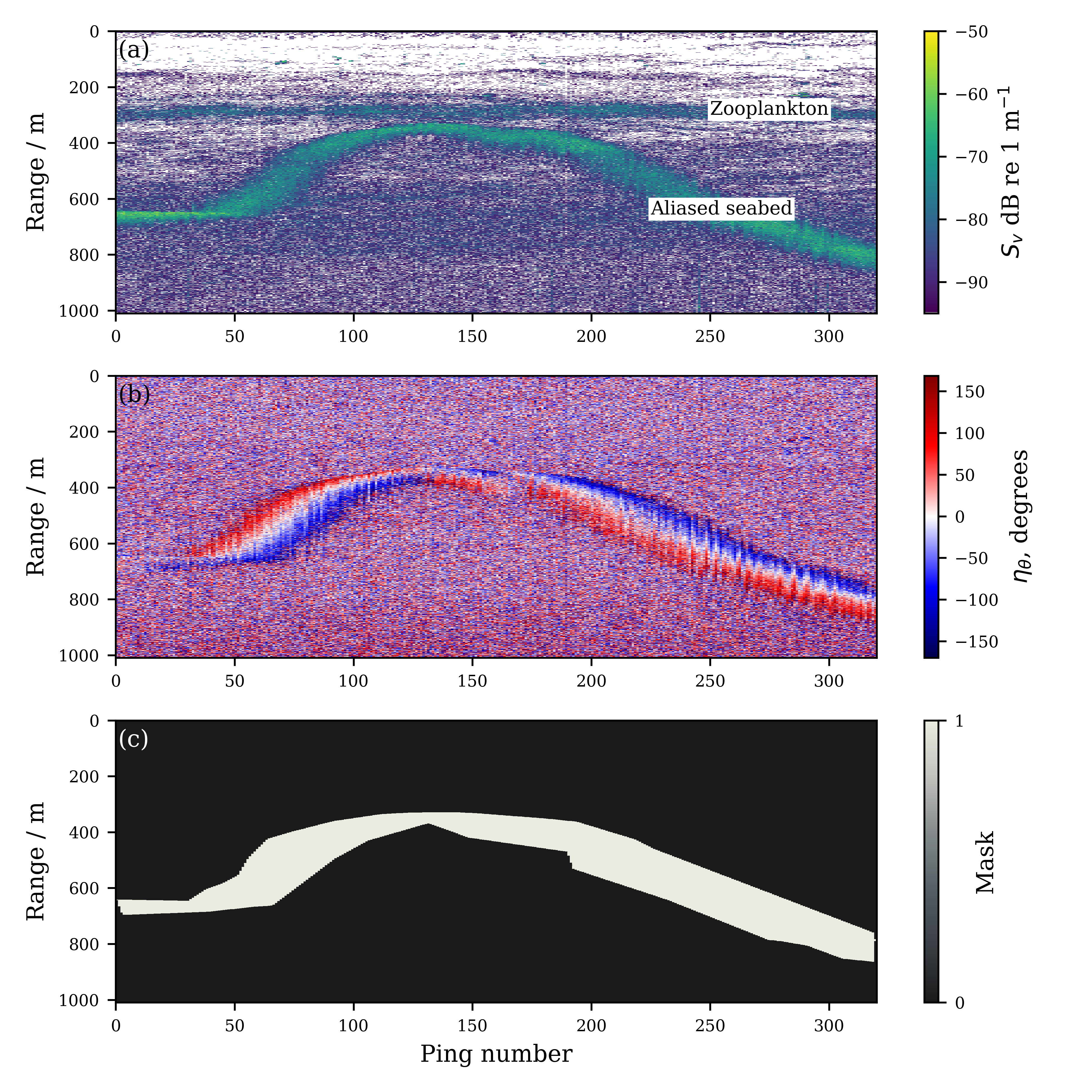}
\caption{Aliased seabed echoes seen in a section of \SI{38}{kHz}
  acoustic data with (a) volume backscatter ($S_v$), (b) along-ship
  split beam angle ($\eta_{\theta}$) and (c) a typical, hand-drawn
  aliased seabed removal mask. The horizontal axis shows pings with
  interval ($I_T$) of \SI{2}{\s}, nominal speed \SI{10}{\knot} and an
  extent of about \SI{3.3}{\km}. Data recorded using a Simrad EK60
  scientific echo sounder on board RRS James Clark Ross, cruise
  JR280.}
\label{fig:example}
\end{figure*}

Echo sounders are routinely used in fisheries acoustics to survey
marine ecosystems \citep{simmonds2008fisheries}. Sound pulses
(``pings'') are transmitted towards a target and the intensity (Volume
backscatter, $S_v$) is measured, integrated and recorded.  Signals in
acoustic data come from a combination of biotic targets (e.g. fish),
abiotic targets (e.g. seabed, gas fluxes) and noise. Therefore,
reflections from biological targets may be obscured by various types
of acoustic noise, corruption or attenuation. Figure
\ref{fig:example}a shows an example echogram where the horizontal
stripe of high $S_v$ is caused by reflections from zooplankton. The
curve of high $S_v$ below is not the seabed, but an alias caused by
seabed reverberations from preceding pings coinciding with the current
ping reception.

Failure to detect and remove unwanted signal prior to biological
target detection could result in poor estimates of animal abundance or
biomass \citep{maclennan2004experiments}. Algorithms exist for the
detection of many of these corruptions: impulsive noise spikes
\citep{anderson2005spatio}; attenuated signal
\citep{ryan2015reducing}; transient noise (persisting for multiple
pings) \citep{ryan2015reducing}; and background noise (relatively
constant for extended periods) \citep{de2007post}. However, aliased
seabed is typically either avoided or removed manually, a notoriously
laborious task. Aliased seabed and biology can have a similar
appearance in echograms (e.g. Figure \ref{fig:example}a), and when
they cross it can be difficult to precisely determine the
boundary. Aliased seabed detection is therefore subjective and a much
harder problem than true seabed detection.

Although aliased seabed can occur at any frequency, it is common in
lower frequency data (e.g. \SI{18}{kHz}, \SI{38}{kHz}) when using a
fixed, short transmit pulse interval ($I_T$) and crossing the
continental shelf. Acoustic signals are attenuated by absorption with
range ($R$) as a function of frequency, temperature and seawater
chemical composition \citep{van2009simple}, limiting echo sounder
range ($R_{max}$). Typical maximum seabed detection depths by
frequency are shown in Table \ref{table:ranges}. If the ping interval
$I_T$ is short with respect to the time taken for a reflection to
occur from a seabed beyond the logging range $R_L$, as described in
Equation \ref{eq1}, then aliasing can occur with reflections from
preceding pings coinciding with echoes from the current ping. However,
practical use of Equation \ref{eq1} for prediction requires detailed
bathymetry data that are rarely available with sufficient spatial
accuracy and resolution (e.g. Global Bathymetric Chart of the Oceans
\citep{ioc2008bodc} $\approx$ \SI{1000}{m} resolution, South Georgia
Bathymetry Database \citep{hogg2016landscape} $\approx$ \SI{100}{m}
resolution) compared to the scale of the acoustic data
(e.g. \SI{10}{m}).  \citet{renfree2016optimizing} present the strategy
for avoiding aliased seabed, by dynamically optimising $I_T$ and the
data logging range ($R_L$).  However, changing parameters mid-survey
causes changes in spatial resolution complicating subsequent data
analysis. In addition, the background noise removal method implemented
by \citet{de2007post} requires a large $R_L$ to determine the noise
level, thus constraining the adjustment demanded by Renfree and Demer.

\begin{table}
    \caption{Maximum seabed detection range ($R_{max}$), using typical transducer
      settings, according to the Simrad EK60 reference manual.}
    \label{table:ranges}
    \begin{center}
      \begin{tabular}{@{}rr@{}}
        \toprule
        Frequency (\si{\kHz}) & $R_{max}$ (\si{\m}) \\
        \midrule
        18 & 7000 \\
        38 & 2800 \\
        70 & 1100 \\
        120 & 850 \\
        200 & 550 \\
        \bottomrule
      \end{tabular}
    \end{center}
\end{table}

\begin{equation}\label{eq1}
R_{\mathcal{A}} = \frac{\mbox{mod}(2 R_S,\  c \ I_T)}{2}, \quad \textrm{where} \quad R_L < R_S < R_{max}
\end{equation}

Modern echo sounders use split-beam transducers, which are divided into
four quadrants allowing target direction to be determined by comparing
the signal received at each quadrant \citep{simmonds2008fisheries}. In
addition to recording amplitude, they also record the split-beam angle
(SBA). The along-ship angle ($\eta_{\theta}$) is the phase difference
between the fore and aft transducer halves, and the athwart-ship angle
($\eta_{\phi}$) is determined from the starboard and port
halves. Reflection and scattering from a deep seabed occur over a
large area due to beam spreading, causing variance in wave arrival
times. A rising seabed is detected at the fore quadrants of the
split-beam transducer before the aft quadrants and vice-versa for a
falling seabed. These effects, caused by the seabed geometry, are
particularly visible in $\eta_{\theta}$ data and appear to
differentiate aliased seabed from biological reflections (Figure
\ref{fig:example}b). \citet{maclennan2004experiments} show that SBA
reflections from fish aggregations are not necessarily an accurate
indication of target direction whilst reflections from the seabed
correlate well to seabed slope.  \citet{bourguignon2009methodological}
show that seabed detection with a Simrad ME70 using SBA and amplitude
together is more effective than using amplitude alone.  This would
suggest that SBA is an additional discriminatory variable.

We present a simple algorithm, based on image processing techniques,
that detects aliased seabed in single-frequency, split-beam acoustic
data, without the need for
\linebreak
bathymetry.

\section{Method}

The patterns seen in SBA are difficult to segment because of
noise. Sample values for $\eta_{\theta}$ and $\eta_{\phi}$ vary
between -128 and 127 (corresponding to \SI{-180}{\degree} to
\SI{180}{\degree} and so we take the mean-squared over a moving window
to smooth the image and accentuate coherent signal (window sizes
determined empirically).  Whilst these pixels fall within the aliased
seabed regions, only a small percentage of area is
identified. However, we can take these pixels and then examine the
surrounding region in $S_v$. Hence, we derive a five-step algorithm:

\begin{enumerate}

\item Find the mean squared of a $28 \times 28$ moving window over
  $\eta_\theta$ and select cells $ > T_{\theta}$ to produce a mask
  $m_1$.

\item Find the mean squared of a $52 \times 52$ moving window over
  $\eta_\phi$ and select cells $ > T_{\phi} $ to produce a mask
  $m_2$.
  
\item Combine the masks $m = m_1 \lor m_2$.

\item Select pixels from $S_v$ using the mask, $m$ and determine the
  median $S_v$ value of the selection to use as a threshold
  $T$. Optionally, $T$ can be constrained to some minimum value
  (e.g. \SI{-70}{dB}).
  
\item Select regions from $S_v$ where $S_v > T$ and which intersect
  $m$. The resulting mask is the union of the selected regions and $m$.
 
\end{enumerate}

We use $T_{\theta} = 702$ and $T_{\phi} = 282$ determined empirically.

The final mask is a grid indicating those pixels that have been
classified as aliased seabed.  It can be used to label aliased seabed
pixels in the original echogram or to replace them, using a suitable
token (e.g. -999) indicating ``no data'' or ``missing value''.

\section{Results}

\begin{figure*}[!htb]
\centering \includegraphics[width=\textwidth]{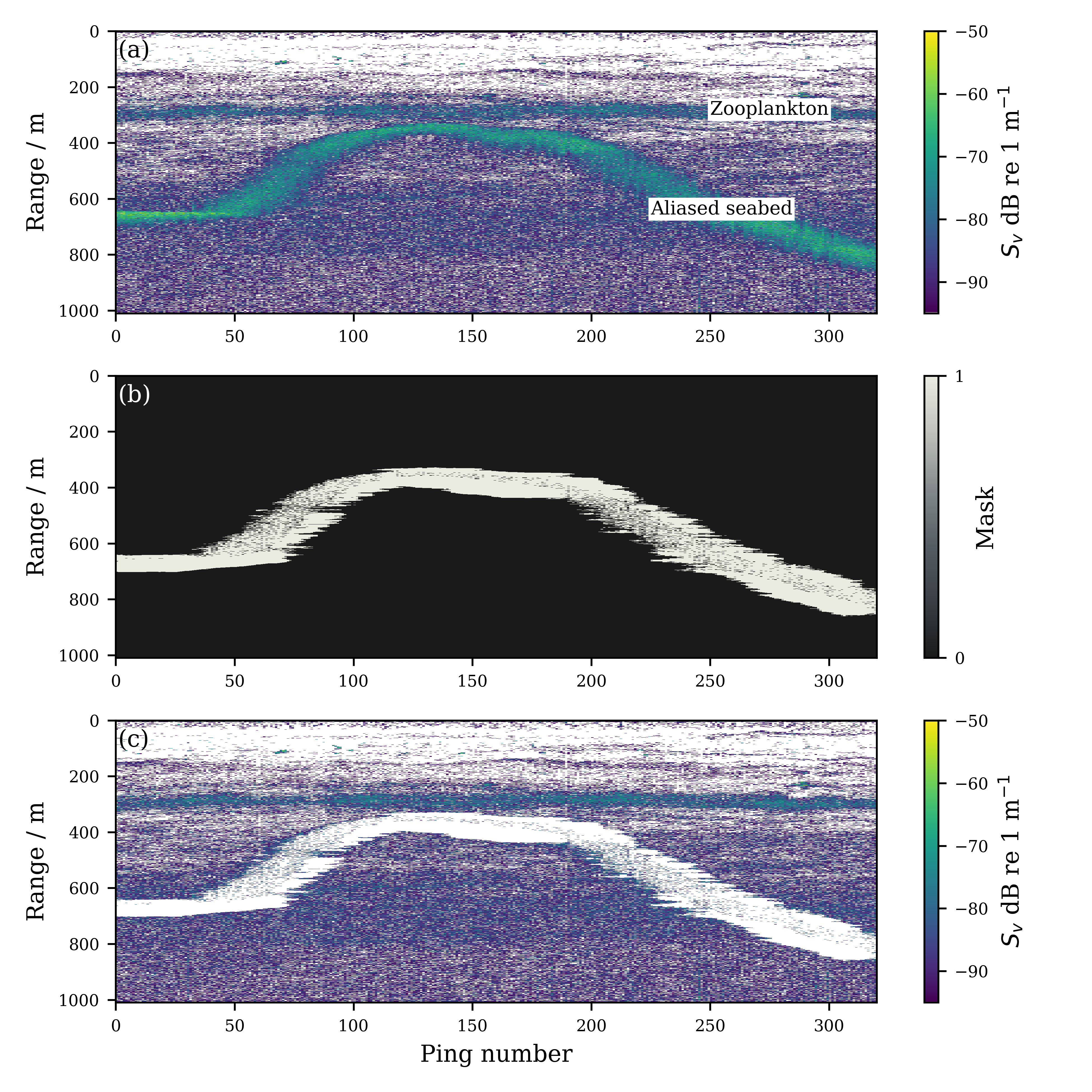}
\caption{Detection and removal of aliased seabed. (a) is the original
  echogram, (b) aliased seabed determined using the algorithm, and (c)
  the echogram with aliased seabed removed.}
\label{fig:results}
\end{figure*}

Figure \ref{fig:results} shows the algorithm applied to our example
data set.

We tested the algorithm on 30 transects from the British Antarctic
Survey annual Western Core Box acoustic survey conducted North West of
South Georgia. Data were collected using a Simrad EK60 echo sounder
(\SI{38}{\kHz}). In all cases, aliased seabed that had been found by
human scrutinization was detected. When $T$ was detected
automatically, we observed six cases of misclassification of
scattering layers as aliased seabed. The misclassification was
eliminated by constraining the threshold $T$ to be at least
\SI{-70}{dB}.

\section{Discussion}

Our algorithm can be used to detect aliased seabed in
single-frequency, split-beam echo sounder data.  The split-beam angle
threshold values, $T_{\theta}$ and $T_{\phi}$, and the convolution
window sizes presented here, are determined empirically from data
collected around South Georgia, where the seabed substrate consists of
fine-grained sediments and clays and there is rapid change in
bathymetry \citep{hogg2016landscape}. The parameters may need to be
adjusted for other vessels and other ocean areas.

Aliased seabed is an additive backscatter corruption, so the algorithm
assumes that the area surrounding an alias is likely to have a lower
backscatter than the alias.  The backscatter threshold ($T$) is
determined dynamically, making the algorithm less sensitive to
calibration correction accuracy. The median is used to determine $T$,
being less susceptible to outliers that the mean.  In practice, it may
be desirable to constrain $T$ to some minimum $S_v$ value to guard
against the possibility of selecting low intensity areas of the
echogram. If $T$ is set too low, then there is a danger that step five
of the algorithm could cause ``leakage'' into surrounding scattering
layers. If $T$ is set too high then the region of aliased seabed
detected is reduced in size. Minimum $T$ should therefore be adjusted
based on observed results.

Speckle is seen in some aliased seabed detections (Figure
\ref{fig:results}b). This can be removed using a hole filling image
processing algorithm (e.g. morphological reconstruction
\citep{soille2013morphological}, as used by Matlab imfill). Visual
inspection of echograms shows that our automated results find several
instances of aliased seabed that had not been identified manually. On
further inspection these appear to be genuine and, in some cases, are
fainter aliases caused by the echoes from antepenultimate pings.

Our algorithm is simple to implement and efficient in terms of
computational resources. The windowing operations can be implemented
using two-dimensional convolution which is fast on modern hardware
(the example in Figure \ref{fig:results} takes about \SI{0.43}{\s} on
a 2016 Intel Skylake i7 processor).  Whilst the algorithm does not
rely on other noise removal strategies beyond true seabed removal, its
performance can reduce if the data include impulse noise, transient
noise or attenuated signal.  In these cases, the methods described by
\cite{anderson2005spatio} and \cite{ryan2015reducing}, combined with
interpolation (e.g. median filtering) to replace the noise, are an
effective preprocessing step. If using background noise removal
(e.g. \cite{de2007post}), we recommend implementing this after aliased
seabed detection.

We want our method to be independent of ping interval ($I_T$) and
logging range ($R_L$), and so we choose to use a single frequency. We
are also interested in using data from ships of opportunity
(e.g. fishing vessels) which may only have a single frequency.
However, If multi-frequency data are available then, depending on
maximum range ($R_{max}$) and seabed depth ($R_S$), other frequencies
can be used to further validate aliased seabed. (E.g. if an aliased
seabed candidate was observed at \SI{500}{\m} in \SI{38}{\kHz} data,
with ping interval $I_T = \SI{2}{\s}$ then, using Equation \ref{eq1},
seabed depth $R_S$ = \SI{2000}{\m}. If a corresponding signal was seen
in \SI{70}{\kHz} data, then the maximum range ($R_{max}$) would be
insufficient to reach the seabed, and so the signal must have another
cause). Lower frequency data could allow $R_S$ to be detected
automatically and allow the methods of \citet{renfree2016optimizing}
to be used as part of a hybrid approach. A consequence of Equation
\ref{eq1} is that $R_S \nless R_L $ and so aliased seabed cannot occur
in a ping where the true seabed has already been detected.

Using split-beam angle in addition to volume backscatter is known to
improve bottom detection
\citep{maclennan2004experiments,bourguignon2009methodological}.  Large
coherent patterns in SBA are a strong indication of reflections from
the seabed, but not biology. We extend this observation to aliased
seabed and use it to create an automated algorithm providing
consistent, repeatable results. We have tested the algorithm with
Simrad EK60 data, which uses a four quadrant, split-beam
configuration.  Some new transducers use a three-sector design,
however we expect the principles to be transferable.  Although we
designed the algorithm for aliased seabed detection, \textit{mutatis
 mutandis}, it may also have applications as a bottom detector.

\section{Conclusions}

The method we describe is intended to make aliased seabed detection
and removal semi-automatic, fast and reproducible. It could be
incorporated into existing tooling to reduce the labour required to
clean fisheries acoustic data. We recommend that practitioners check
results using visual inspection.

\begin{notes}[Acknowledgements]

Our thanks to the officers, crew and scientists onboard the RRS James
Clark Ross and RRS Discovery for their assistance in collecting the
data.

Alejandro Ariza (British Antarctic Survey) provided helpful insight
into the problems caused by aliased seabed.

The Western Core Box cruises and SF are funded as part of the
Ecosystems Programme at the British Antarctic Survey, Natural
Environment Research Council, a part of UK Research and Innovation.

This work was supported by the Natural Environment Research Council
grant
\linebreak
{NE/N012070/1}.

\end{notes}

\end{document}